\documentclass[conference]{IEEEtran}
\IEEEoverridecommandlockouts

\usepackage{cite}
\usepackage{caption}
\usepackage{amsmath,amssymb,amsfonts}
\usepackage{algorithmic}
\usepackage{graphicx}
\usepackage{textcomp}
\usepackage{makecell}
\usepackage{url}
\usepackage{xcolor}
\usepackage{tcolorbox}
\usepackage{balance}

\def\BibTeX{{\rm B\kern-.05em{\sc i\kern-.025em b}\kern-.08em
    T\kern-.1667em\lower.7ex\hbox{E}\kern-.125emX}}

\usepackage{listings}
\makeatletter
\def\lst@makecaption{%
  \def\@captype{table}%
  \@makecaption
}

\newtcolorbox{resultbox}{colback=gray, arc=0.5mm, top=1mm, bottom=1mm, left=1mm, right=1mm}
\newtcolorbox{promptbox}{colback=white, arc=0.5mm, top=1mm, bottom=1mm, left=1mm, right=1mm, title=Prompt Templates used for refactoring}

\makeatother
\begin{document}

\title{Automated Refactoring of Non-Idiomatic Python Code: A Differentiated Replication with LLMs}

\author{
\IEEEauthorblockN{Alessandro Midolo}
\IEEEauthorblockA{\textit{Department of Engineering} \\
\textit{University of Sannio}\\
Benevento, Italy \\
amidolo@unisannio.it}
\and
\IEEEauthorblockN{Massimiliano Di Penta}
\IEEEauthorblockA{\textit{Department of Engineering} \\
\textit{University of Sannio}\\
Benevento, Italy \\
dipenta@unisannio.it}
}

\maketitle
\thispagestyle{plain}
\begin{abstract}
In the Python ecosystem, the adoption of idiomatic constructs has been fostered because of their expressiveness, increasing productivity and even efficiency, despite controversial arguments concerning familiarity or understandability issues. Recent research contributions have proposed approaches---based on static code analysis and transformation---to automatically identify and enact refactoring opportunities of non-idiomatic code into idiomatic ones.
Given the potential recently offered by Large Language Models (LLMs) for code-related tasks, in this paper, we present the results of a replication study in which we investigate GPT-4 effectiveness in recommending and suggesting idiomatic refactoring actions.  Our results reveal that GPT-4 not only identifies idiomatic constructs effectively but frequently exceeds the benchmark in proposing refactoring actions where the existing baseline failed. A manual analysis of a random sample shows the correctness of the obtained recommendations. Our findings underscore the potential of LLMs to achieve tasks where, in the past, implementing recommenders based on complex code analyses was required.
\end{abstract}


\section{Introduction}

Pythonic idioms, defined by conventions and practices adhering to the principles and style of the language, are a hallmark of Python's expressive power. Adopting Pythonic idioms is seen as essential for ensuring that Python code remains approachable for both beginners and seasoned developers, promoting consistency across diverse codebases and encouraging the use of Python’s expressive features~\cite{pep2000developers, carol2018usage}.
Several studies have explored Pythonic idioms from various perspectives. These include the prevalence of such idioms \cite{carol2018usage}, their understandability ~\cite{zid2024study}, and fix-inducing proneness\cite{ZampettiBZAP22}.

To ease refactoring toward Pythonic idioms, Zhang et al.~\cite{zhang2024automated, zhang2022making, zhang2023ridiom} proposed a static analysis tool to identify and refactor anti-idiomatic code smells, or non-idiomatic code. 

The advent of Large Language Models (LLMs) has significantly influenced software engineering, showcasing remarkable capabilities in generating code, assisting in debugging and solving complex programming challenges~\cite{fan2023large, wei2023copiloting}. Notably, LLMs have shown the potential to outperform human developers in identifying code refactoring opportunities~\cite{cordeiro2024empirical}.

Motivated by the increasing adoption of LLMs in software engineering tasks and the growing scholarly interest in Pythonic idioms, this paper replicates the work of Zhang et al.\cite{zhang2024automated} using GPT-4. Our study evaluates GPT-4's performance in refactoring anti-idiomatic code smells into Pythonic idioms. We leverage the benchmark dataset introduced in\cite{zhang2024automated}, comprising 1,061 Python methods extracted from 736 public GitHub repositories, which highlight non-idiomatic code snippets. The benchmark, curated manually by the authors and 24 developers, serves as the baseline for comparison. Additionally, we manually assess 120 randomly selected methods from the dataset to evaluate the correctness of refactorings proposed by GPT-4 and the benchmark.

Our results indicate that GPT-4  outperforms the benchmark, achieving a 28.8\% increase in the number of refactoring proposals. While GPT-4 aligns with the benchmark in many cases, it also introduces novel refactorings not captured by the original dataset, showcasing its broader applicability. However, GPT-4 occasionally struggles with complex or nuanced code patterns, highlighting areas for potential improvement. In terms of correctness, GPT-4 achieves a 90.7\% accuracy compared to the benchmark’s 76.8\%, reflecting its ability to generate more accurate and reliable refactorings overall. These results indicate that GPT-4 not only identifies more refactorings but also does so with greater precision, though there is room for enhancement in handling edge cases.

The study replication package is publicly available~\cite{GPTIdiomRefactoring}.

\section{Related Work} \label{sec:related}
This section discusses related work about (i) studies on Pythonic Idioms, (ii) Pythonic idiom and functional constructs' refactoring, and (iii) the use of LLM in software engineering. 

\subsection{Pythonic Idioms}
Python is a very versatile language, allowing one to write source code using different paradigms, including procedural, object-oriented, and functional. One peculiar aspect of Python is in its idioms~\cite{pep2000developers}. Many books and online materials~\cite{carol2018usage, merchante2017python, slatkin2020effective} promote the use of Pythonic idioms for their concise coding styles. 

Zid et al.~\cite{zid2024list} investigate the performance impact of using list comprehensions, a Pythonic functional construct for concise list processing, compared to equivalent imperative for-loop constructs. Their study systematically evaluates their performance using a set of transformation rules to convert for loops into list comprehensions. Testing artificial code snippets reveals that list comprehensions are generally faster than procedural code, with pronounced performance differences when amplifying test repetitions. Also, Zid et al.~\cite{zid2024study} explore the impact of Pythonic functional constructs—such as lambdas, comprehensions, and map/reduce/filter functions—on code understandability compared to procedural alternatives. Through a controlled experiment, they found that 
procedural code is easier to modify than complex lambdas, while complex comprehensions
are easier to modify than their procedural counterparts.
In general, participants perceived functional constructs as more difficult to understand than their procedural counterparts. In another study, Zampetti et al. \cite{ZampettiBZAP22} found that Pythonic functional constructs have higher chances of inducing fixes than other changes.

Carol et. al~\cite{carol2018usage} explore Pythonic idioms in Python programming, discussing how these idioms contribute to the language’s style and culture. It also examines how Python developers, especially experienced ones, understand the concept of "Pythonic" code. They highlight the importance of using idiomatic constructs in Python for cleaner, more readable, and maintainable code. Sakulniwat et al~\cite{sakulniwat2019visualizing} investigates the evolution of Pythonic idioms, particularly focusing on the 'with open' idiom for file handling in Python projects. Leelaprute et al~\cite{leelaprute2022coding} explores the performance impact of using Pythonic idioms in large-scale data processing. Their findings indicate that Pythonic code can indeed offer performance improvements, especially in terms of memory usage and execution time. Similarly, Zhang et al.~\cite{zhang2023faster} investigates the performance implications of various Pythonic idioms. Their study aims to address the conflicting views in the community regarding whether idiomatic Python code performs better or worse than its non-idiomatic counterparts. The results show that the performance impact of idiomatic code is not always straightforward. Factors such as variable scope, the complexity of the idiom, and the execution paths involved can all influence whether an idiomatic expression performs better or worse.

\subsection{Automated Code Refactoring}
By minimizing human intervention, automated refactoring reduces the risk of introducing errors while ensuring consistent adherence to best practices, making it an essential component in modern software development workflows.

\begin{table*}[ht!]
    \caption{Description of the twelve Pythonic idioms used to refactor non-idiomatic Python code}
    \begin{center}
        \begin{tabular}{|l|l|l|}
            \hline
            \textbf{Name} & \textbf{pep\_ref} & \textbf{Description} \\
            \hline
            list comprehension & PEP 202 & \makecell[l]{introduce a concise syntax for creating lists by applying an expression to each element\\ in an iterable, with optional filtering} \\
            \hline
            for multiple targets & & \makecell[l]{allows unpacking multiple values into separate variables within a loop. This enables\\ iteration over iterable objects like lists or tuples, assigning each value to the\\ corresponding target variable in a single step} \\
            \hline
            chain comparison &  & \makecell[l]{allow multiple relational conditions in a single expression. It’s useful for range checks\\ and ordered comparisons} \\
            \hline
            fstring & PEP 498 & \makecell[l]{introduces a more efficient and readable way to format strings. It allows expressions\\ to be embedded directly inside string literals. This simplifies string interpolation by \\evaluating expressions at runtime and directly embedding their results within the string} \\
            \hline
            assign multiple target & & \makecell[l]{allow assigning multiple values in one statement. If consecutive assignments don't\\ depend on each other, they can be refactored into a single assignment} \\
            \hline
            star in function calls & PEP 448 & \makecell[l]{introduces enhanced unpacking syntax in Python, allowing more flexible and concise\\ unpacking of iterables and dictionaries. It supports extended unpacking in lists and\\ merging dictionaries with the ** operator} \\
            \hline
            truth value test & & \makecell[l]{check an object's truthiness without needing explicit comparisons. Objects like None,\\ zero, and empty lists are inherently false. Non-idiomatic code arises when comparisons \\to these are made directly, though not all cases allow refactoring} \\
            \hline
            loop else & & \makecell[l]{the else clause of a loop is executed only if the loop finishes without a break statement\\ interrupting it. If the loop ends due to a break, the else block is not executed} \\
            \hline
            with & PEP 343 & \makecell[l]{introduce the with statement for context management in Python. It defines a standard for\\ managing resources like files or locks, ensuring they are properly acquired and released}\\
            \hline
            dict comprehension & PEP 274 & \makecell[l]{offer a concise way to create dictionaries by defining key-value pairs through an \\expression, optionally with filtering conditions}\\
            \hline
            chain assign same value & & \makecell[l]{involve assigning the same value to multiple variables in a single expression. It leverages\\ the ability to chain assignments, where each variable on the left-hand side receives the\\ same value in one operation} \\
            \hline
            set comprehension &  & \makecell[l]{provide a concise way to construct sets by applying an expression to each element in \\an iterable, with optional filtering} \\
            \hline
        \end{tabular}
        \label{tab:idioms}
    \end{center}
\end{table*}

Zhang et al.~\cite{zhang2023ridiom, zhang2024automated} propose RIdiom, an automatic refactoring tool that transforms non-idiomatic Python code into idiomatic equivalents by identifying Pythonic idioms through a systematic comparison of Python and Java syntax, defining syntactic patterns for detecting non-idiomatic code, and employing AST-based rewriting operations. Despite the high precision achieved, the tool exhibits a low recall (the lowest 0.66 in list comprehension idiom) on several refactorings, missing the opportunity to refactor a large amount of anti-idiomatic code smells. Also, Zhang et al.~\cite{zhang2024hard} tackle the challenge of understanding Pythonic idioms in programming. It introduces DeIdiom, a tool designed to convert Pythonic idioms into more verbose, non-idiomatic code to improve comprehension, especially for developers unfamiliar with these idioms. The study highlights how idiomatic code, while concise, can be hard to understand for some readers, particularly when they are new to Python or come from different programming languages.

The introduction of Java 8 brought functional programming into the Java ecosystem, enabling concise, expressive code through features like streams and lambda expressions while reducing common errors~\cite{urma2018modern}. Several approaches have since emerged to automate the refactoring of for loops into stream pipelines. LambdaFicator~\cite{franklin2013lambdaficator} automates tasks such as converting anonymous inner classes to lambda expressions and transforming for loops into stream operations like map and filter~\cite{gyori2013crossing}. Similarly, other methods use preconditions and templates to identify loops suitable for refactoring, replacing them with equivalent stream-based code while preserving functionality~\cite{midolo2021refactoring}.

Khatchadourian et al.\cite{khatchadourian2020safe} focus on analyzing stream operations for safe parallelization, defining conditions to ensure performance gains when converting sequential to parallel streams. Kayak\cite{david2017kayak}, a semantics-driven refactoring tool, utilizes program synthesis to transform external iterations into Java 8 Streams.  Tsantalis et al.~\cite{tsantalis2017clone} explore using lambda expressions to refactor behavioral software clones, addressing structural and functional variations in Type-2 and Type-3 clones, demonstrating the effectiveness of their approach on extensive datasets.

\subsection{LLMs for Software Engineering}

Several studies investigated the application of LLMs. to software engineering problems. In the following, we will discuss some relevant work. A more extensive analysis is available in systematic literature reviews \cite{xinyi2024large, fan2023large}. 

Cordeiro et al.~\cite{cordeiro2024empirical} demonstrate how StarCoder2, a large language model (LLM), can enhance software refactoring by automating systematic code improvements while reducing code smells. The study finds that StarCoder2 is effective in repetitive, rule-based refactoring tasks, whereas developers are better at addressing complex, context-dependent issues. Depalma et al.\cite{depalma2024exploring} assess its Java code refactoring skills, focusing on optimizing loops and improving code quality. White et al.\cite{white2024chatgpt} propose prompt patterns to tackle challenges like decoupling code from libraries and generating API specifications. Chavan et al.~\cite{chavan2024analyzing} analyze developer interactions with ChatGPT on GitHub and Hacker News, identifying its role in code refactoring.

Guo et al.\cite{guo2024exploring} evaluate ChatGPT's performance in automated code review, showing it outperforms CodeReviewer on benchmarks. Siddiq et al.\cite{siddiq2024quality} use the DevGPT dataset to investigate the quality and use of ChatGPT-generated Python and Java code, highlighting issues like security flaws and poor documentation.  
Tufano et al.~\cite{tufano2024unveiling} mine ChatGPT mentions on GitHub, and leverage them to create a taxonomy of 45 developer tasks automated with ChatGPT. Unsurprisingly, these tasks include refactoring. Poldrack et al.~\cite{poldrack2023ai} explores the capabilities of GPT-4 to refactor code improving quality metrics such as maintainability, cyclomatic complexity, and compliance with coding standards. Noever et al.~\cite{noever2023chatbots} investigates how AI-driven code assistants can analyze and improve historically significant pieces of code. It highlights the role of AI in providing insights into obfuscated or poorly documented code, improving clarity, and enhancing performance.

\section{Study Design} \label{sec:design}
The \emph{goal} of our study is to investigate the performance of LLMs, specifically GPT-4, in refactoring Python source code to adopt Pythonic idioms. 
The \emph{quality focus} is the ability of the LLM to automatically refactor Python code towards Pythonic idioms, as compared to alternative approaches based on code analysis and transformation, and predefined heuristics and patterns\cite{zhang2024automated,zhang2022making}.
By doing so, we aim to understand how a language model such as GPT-4 performs in comparison to traditional, rule-based methods that rely on a set of deterministic procedures for code transformation.

The study  addresses three research questions:
\begin{itemize}
    \item \textbf{RQ1}: \textit{How does GPT-4 perform in refactoring non-idiomatic Python code compared to the state-of-the-art (SOTA) benchmark?} This question aims to assess the feasibility of GPT-4's refactorings in comparison with a well-established deterministic approach. The comparison with the SOTA benchmark will provide insights into how GPT-4, as a generative model, can potentially offer advantages over traditional deterministic approaches.
    \item \textbf{RQ2}: \textit{In which specific cases does GPT-4 fail to successfully perform a refactoring action, and what underlying factors contribute to these failures?} This question seeks to explore the limitations of GPT-4 in terms of understanding and applying Pythonic idioms, as well as any challenges that may arise in particular code contexts.
    \item \textbf{RQ3}: \textit{To what extent are the refactoring actions proposed by GPT-4 correct?} This question examines whether GPT-4's transformations maintain functional equivalence, a critical aspect of reliable and trustworthy code refactoring.
\end{itemize}

By investigating these questions, this study aims to provide a detailed assessment of GPT-4's strengths and limitations in automated code refactoring, contributing to the broader discourse on the applicability of large language models in software engineering tasks. All data analyzed, including the refactoring outputs generated by GPT-4, along with the associated metrics and results, are available in the paper's public GitHub repository~\cite{GPTIdiomRefactoring}.

\subsection{Dataset and Baseline}
As a dataset (and baseline) for our study, we adopt the benchmark proposed by Zhang et al.~\cite{zhang2024automated, pythonicIdiom}, which produced a dataset consisting of 1,061 Python methods extracted from 736 public repositories on GitHub. This dataset provides a comprehensive collection of code samples, reflecting a wide range of coding practices across different Python projects. To ensure consistency with the existing benchmark, we focus on the twelve Pythonic idioms identified and discussed by Zhang et al. ~\cite{zhang2024automated}. Table~\ref{tab:idioms} reports a summary of these Pythonic idioms, which have been selected based on their frequency and impact on code readability and efficiency. The idioms are associated with a reference number from the Python Enhancement Proposals (PEP) list~\cite{pep2000developers}, although not all idioms are formally indexed within the PEP system. The benchmark dataset includes 1,814 pairs of code snippets, each consisting of a "non-idiomatic" code snippet and its refactored version that incorporates one of these twelve idioms.

The benchmark provides the original method from which the non-idiomatic code was identified, along with the non-idiomatic code itself and its refactored counterpart. During our analysis, we encountered a few issues with the dataset. Specifically, some of the original methods referenced in the dataset were missing, which required us to intervene and source the missing data to ensure that all necessary components were available for the refactoring process. Additionally, we identified and removed duplicate entries within the dataset to maintain the integrity of the benchmark.

\subsection{GPT-4 generation process}

The next phase of our study focuses on refactoring methods containing anti-idiomatic code smells into idiomatic Python code using GPT-4. This process is automated through the OpenAI API, which facilitates programmatic interaction with the model. For each of the twelve selected Python idioms, we run the generation process for all methods in the benchmark associated with that idiom. The specific model used for the generation is \textit{chatgpt-4o-latest}(i.e. gpt-4o-2024-08-06). with default parameters (temperature=1.0, top\_p=1.0, max\_tokens=4000).

To ensure consistent generation of Pythonic code, we conducted manual experiments in prompt engineering to identify the optimal prompt for setting the context of the desired refactoring action. We observed that explicitly specifying the Pythonic idiom and the programming language, both in the context and in the prompt provided to the LLM, helps GPT-4 focus on the target idiom and avoid applying unrelated refactoring actions.

We set the context with the "\textit{system}" prompt in the following textbox, where [pythonic\_idiom] represents the specific Python idiom being targeted. This context encourages the model to align its output with modern Python best practices.

Each refactoring task is triggered by a prompt that combines the selected idiom and the relevant code ("\textit{user}" prompt in the textbox). We also request the model to count the number of refactorings performed, as the input method may contain multiple instances of anti-idiomatic patterns related to the same Python idiom. This count allows us to compare the number of refactorings suggested by the model with those proposed by the benchmark.

For Python idioms associated with formal PEPs, we reference the corresponding PEP in the context and the generation prompt (e.g., list comprehensions are referenced as "PEP 202 - List Comprehensions"). For idioms without formal PEP definitions, we use their extended names (e.g., "chain\_ass" is referred to as "chain assignment of the same value"). In cases where the idioms lack clear definitions, we conduct preliminary tests to ensure that GPT-4 accurately understands and applies the intended idiomatic pattern.

\begin{promptbox}
\textbf{System}: "You are a software developer, skilled in writing Python source code and refactoring Python code using [pythonic\_idiom]"

\textbf{User}: "Refactor the code using the [pythonic\_idiom] idiom, and provide the refactored Python code along with the number of [pythonic\_idiom] refactorings you have made: [code]"
\end{promptbox}

\subsection{Analysis Methodology}
To address RQ1 and RQ2, we conduct a comparative analysis of the refactorings generated by GPT-4 and those documented in the benchmark. The first step involves summarizing the total number of refactorings performed for each Pythonic idiom, providing a high-level comparison of the two approaches. This initial overview helps identify trends and differences in the overall refactoring coverage. Next, we examine specific instances where the number of refactorings performed per method differs, categorizing cases where GPT-4 generates either a greater or smaller number of refactorings compared to the benchmark. This analysis highlights GPT-4's relative strength or limitations in addressing different idioms. For RQ2, we focus on failure scenarios where GPT-4 does not generate any refactorings or produces fewer than those proposed in the benchmark. This step is critical for understanding GPT-4's gaps in effectively handling certain idioms.

\begin{table*}[t]
    \caption{Summary of refactoring metrics comparing GPT-4 and the benchmark~\cite{zhang2024automated} for various Pythonic idioms. Metrics include total methods analyzed, refactorings proposed by each approach, agreement cases, and instances where one approach outperformed the other}
    \begin{center}
        \begin{tabular}{|l|r|r|r|r|r|r|r|}
            \hline
            \textbf{pythonic idiom} & \textbf{methods} & \textbf{count\_gpt} & \textbf{count\_bench} & \textbf{equals} & \textbf{gpt\_more} & \textbf{bench\_more} & \textbf{gpt\_zero} \\
            \hline
            list comprehension & 100 & 205 & 131 & 63 & 36 & 1 & 0\\
            \hline
            for multiple targets & 100 & 197 & 110 & 63 & 37 & 0 & 0 \\
            \hline
            chain comparison & 100 & 166 & 132 & 56 & 32 & 12 & 6 \\
            \hline
            fstring & 100 & 285 & 225 & 76 & 23 & 1 & 0 \\
            \hline
            assign multiple target & 100 & 176 & 157 & 55 & 26 & 18 & 1 \\
            \hline
            star in function calls & 100 & 145 & 127 & 67 & 21 & 12 & 2 \\
            \hline
            truth value test & 100 & 287 & 161 & 43 & 54 & 3 & 1 \\
            \hline
            loop else  & 100 & 129 & 113 & 78 & 15 & 6 & 0\\
            \hline
            with & 100 & 155 & 132 & 82 & 14 & 4 & 0 \\
            \hline
            dict comprehension & 101 & 157 & 120 & 68 & 31 & 2 & 0\\
            \hline
            chain assign same value & 100 & 223 & 227 & 58 & 25 & 17 & 1 \\
            \hline
            set comprehension & 100 & 140 & 123 & 78 & 18 & 4 & 0 \\
            \hline
            \hline
            \textbf{total} & 1201 & 2265 & 1758 & 787 & 332 & 80 & 11 \\
            \hline
        \end{tabular}
        \label{tab:metrics}
    \end{center}
\end{table*}

\begin{figure*}[t!]
\centering
\noindent\begin{minipage}{.45\textwidth}
\begin{lstlisting}
def test_send_video_formatting_caption(self):
    file_data=open('./test_data/test_video.mp4', 'rb')
    tb = telebot.TeleBot(TOKEN)
    ret_msg = tb.send_video(CHAT_ID, file_data,
        caption='_italic_', parse_mode='Markdown')
    assert ret_msg.caption_entities[0].type=='italic'
\end{lstlisting}
\end{minipage}
\hspace{10pt}
\begin{minipage}{.45\textwidth}
\begin{lstlisting}
def test_send_video_formatting_caption(self):
    with open('./test_data/test_video.mp4', 'rb')
        as file_data:
        tb = telebot.TeleBot(TOKEN)
        ret_msg = tb.send_video(CHAT_ID, file_data,
            caption='_italic_', parse_mode='Markdown')
        assert ret_msg.caption_entities[0].type ==
            'italic'
\end{lstlisting}
\end{minipage}
\captionof{lstlisting}{An example of refactoring generated by GPT-4, where the 'with' idiom is used to handle the opening of a mp4 file. On the left side, the original code extracted from the repository. On the right side, the refactoring proposed by GPT-4.}
\label{lst:equal}
\end{figure*}

To address RQ3, we conduct a deeper qualitative evaluation by randomly selecting 120 methods from the twelve Pythonic idioms. The selection process ensures proportional representation, with the number of sampled methods for each idiom reflecting the idiom's contribution to the total refactorings. Each selected refactoring is manually inspected to assess its correctness in two dimensions: adherence to Python syntax and preservation of the original code behavior. For each method selected, we compare the original code and the refactored version, both for benchmark and GPT-4. First, we identify the refactored portion of the source code, and then we check for syntactic errors that could prevent compilation. If no syntactic issues are found, we focus on comparing the logic of the original and refactored code to ensure the behavior has been preserved. The goal of this manual validation is to check that refactorings are not only syntactically valid but also semantically equivalent to the original implementation, thereby guaranteeing functional correctness.

\section{Results} \label{sec:results}
This section presents the outcomes of the study, presenting both quantitative and qualitative results of Pythonic idioms generated by GPT-4 and the benchmark~\cite{zhang2024automated}.

Table~\ref{tab:metrics} reports metrics related to the refactorings proposed by GPT-4 and the baseline benchmark. Column \textit{methods} refers to the total number of methods analyzed, column \textit{count\_gpt} shows the number of refactorings proposed by GPT-4, while \textit{count\_bench} indicates the number of refactorings in the benchmark. The \textit{equals} column denotes the number of methods where both approaches suggest an equal number of refactorings. The \textit{gpt\_more} column reports cases in which GPT proposes more refactorings, whereas \textit{bench\_more} reflects cases where the benchmark suggests more refactorings. Finally, the \textit{gpt\_zero} column represents the instances where GPT-4 provides no refactorings.

In the following, we report and discuss the results addressing the three research questions formulated in Section \ref{sec:design}.

\subsection{RQ1: GPT-4 baseline performance}

Across the dataset, GPT-4 proposes a total of 2265 refactorings, a 28.8\% increase compared to the benchmark’s 1758 refactorings. This significant margin highlights GPT-4's advanced ability to identify potential improvements in Python code. Notably, idioms such as \textit{fstring} and \textit{truth\_test} showcase GPT-4’s performance, with 285 and 287 refactorings, respectively, compared to 225 and 161 by the benchmark.

Despite GPT-4’s dominance in proposing refactorings, there is a substantial overlap between its outputs and those of the baseline benchmark. The \textit{equals} column indicates an agreement in 787 methods, representing over 65\% of the total methods analyzed. The agreement is defined as both GPT-4 and the benchmark suggesting the same number of refactorings for a given method. A textual match was not performed, as two refactorings may differ semantically while remaining syntactically consistent.

\begin{figure*}[t!]
\centering
\noindent\begin{minipage}{.45\textwidth}
\begin{lstlisting}
def test_agent_fit_setter():
    new_agent = agent.Agent(1, 1, 0, 1)
    new_agent.fit = np.array([0])
    if not new_agent.fit.any():
# use truth value to check for an array of zeros
        new_agent.fit = 0
    assert not new_agent.fit
# use direct truthiness check instead of `== 0`
\end{lstlisting}
\end{minipage}
\hspace{10pt}
\begin{minipage}{.45\textwidth}
\begin{lstlisting}
def test_agent_fit_setter():
    new_agent = agent.Agent(1, 1, 0, 1)
    try:
        new_agent.fit = np.array([0])
    except:
        new_agent.fit = 0
    assert not new_agent.fit 
#before the refactoring: new_agent.fit == 0
\end{lstlisting}
\end{minipage}
\captionof{lstlisting}{An example of refactoring where GPT-4 proposes more refactoring actions than the benchmark~\cite{zhang2024automated}. The Pythonic idiom used in this refactoring is the \textit{truth value test}. On the left side, the refactoring proposed by GPT-4. On the right side, the refactoring proposed in the benchmark.}
\label{lst:more}
\vspace{-2mm}
\end{figure*}

Idioms such as \textit{loop\_else} and \textit{with} showcase the highest levels of agreement, with 78 and 82 methods, respectively, receiving identical refactoring counts from both approaches. These cases of alignment are particularly notable for idioms that involve syntactic constructs with less ambiguity, where the application of refactoring is more straightforward.

From a practical standpoint, this high level of agreement also has implications for developers and researchers. It suggests that incorporating GPT-4 into existing workflows would not disrupt the foundational insights provided by benchmark tools but would rather enhance them, ensuring a balance between innovation and reliability.

Listing~\ref{lst:equal} shows an example of refactoring proposed by GPT-4. The method is present in the benchmark as part of the refactoring using the \textit{with} pythonic idiom. The code is taken from the file \textit{test\_telebot.py} from the \textit{pyTelegramBotAPI} repository on GitHub~\cite{pyTelegramBotAPI}. On the left side, the figure shows the original code extracted from the repository, while on the right side, it shows the refactored version proposed by GPT-4.

In the original implementation, the file is opened using the standard \texttt{open} function without an explicit mechanism to ensure its closure after use. The proposed refactoring introduces a \textit{with} statement, which implicitly handles the closing of the file once the code block is executed. This approach not only eliminates the need for a separate \texttt{close} call but also safeguards against potential issues arising from open file handles in cases of exceptions or premature exits from the block. The refactoring is noteworthy for its alignment with the benchmark’s proposed improvements emphasizing the code’s reliability and adhering to Python best practices for resource management.

The \textit{gpt\_more} column reveals that GPT-4 identifies additional refactoring opportunities in 332 methods, which is a remarkable 4-to-1 ratio compared to the 80 methods where the benchmark proposes more refactorings (column \textit{bench\_more}). This disparity underscores GPT-4's superior capability in recognizing a broader range of refactoring possibilities, making it a valuable tool for comprehensive code improvement.

Such differences are particularly pronounced for idioms such as \textit{truth value test} and \textit{dict comprehension}, where GPT-4 consistently outperforms the benchmark in proposing additional refactorings. For \textit{truth value test}, GPT-4 identifies 54 methods with more refactorings compared to the benchmark, reflecting its nuanced capability to optimize logical conditions and eliminate unnecessary redundancy in code. Similarly, for \textit{dict comprehension}, GPT-4 proposes additional improvements in 31 methods, highlighting its adeptness at recognizing opportunities to enhance dictionary creation through Pythonic constructs. Such performance suggests that GPT-4 excels in areas requiring a deeper syntactic and semantic understanding of Python code. Its ability to uncover sophisticated improvements reflects not only the richness of its training but also its capacity to generalize across diverse coding patterns. 

Listing~\ref{lst:more} shows a comparison between the refactoring proposed by GPT-4 and the refactoring proposed in the benchmark. In this case, GPT-4 proposed two \textit{truth value test} refactorings compared to the one proposed in the benchmark. The code is taken from the file \textit{test\_agent.py} from the \textit{opytimizer} repository on GitHub~\cite{opytimizer}.

\begin{figure*}[t!]
\centering
\noindent\begin{minipage}{.45\textwidth}
\begin{lstlisting}
def test_frame_selector_random_k_2(self):
    _SEED, _K = 43, 10  # Combined assignments
    random.seed(_SEED)
    selector = RandomKFramesSelector(_K)
    frame_tss = list(range(0, 6, 2))
    _SELECTED_GT = [0, 2, 4]
    selected = selector(frame_tss)
    self.assertEqual(_SELECTED_GT, selected)
\end{lstlisting}
\end{minipage}
\hspace{10pt}
\begin{minipage}{.45\textwidth}
\begin{lstlisting}
def test_frame_selector_random_k_2(self):
    _SEED , _K  = 43, 10 # before split
    random.seed(_SEED)
    selector , frame_tss  = #before split
        RandomKFramesSelector(_K), list(range(0, 6, 2))
    _SELECTED_GT , selected  = #before split
        [0, 2, 4], selector(frame_tss)
    self.assertEqual(_SELECTED_GT, selected)
\end{lstlisting}
\end{minipage}
\captionof{lstlisting}{An example of refactoring where the benchmark~\cite{zhang2024automated} proposes more refactoring actions than GPT-4. The Pythonic idiom used in this code is the \textit{assign multiple target}. On the left side, the refactoring proposed by GPT-4. On the right side, the refactoring proposed in the benchmark.}
\label{lst:less}
\end{figure*}

The first refactoring is the one proposed by GPT-4, yet omitted by the benchmark. It addresses the issue in the original code where the line \texttt{new\_agent.fit = np.array([0])} could lead to an exception under certain conditions. Although the specific error is not clearly defined, the intention likely was to catch cases where \texttt{new\_agent.fit} was improperly set. The original code resolves this by using a \texttt{try-except} block, which catches the potential exception and resets \texttt{new\_agent.fit} to 0. While this approach is functional, it introduces the overhead of exception handling, creating a trade-off between improved robustness and added complexity.

In contrast, GPT-4 proposes a more Pythonic approach, replacing the exception-based handling with the condition \texttt{if not new\_agent.fit.any()}. The \texttt{.any()} method checks if any element in the \emph{numpy} array is equal to \emph{True}. If all elements of the array are zero or \emph{False}, the condition evaluates as \emph{False}, triggering the assignment of 0 to \texttt{new\_agent.fit}. This method removes the reliance on exceptions and instead uses explicit truth-value testing, a clearer and more efficient solution.

The second refactoring suggested by GPT-4 aligns with the refactoring proposed in the benchmark. In the original code, the assertion \texttt{assert new\_agent.fit == 0} is used to verify that \texttt{new\_agent.fit} equals zero. GPT-4 refactors this to \texttt{assert not new\_agent.fit}, which more directly utilizes Python’s truthiness semantics. In this refactoring, Python evaluates \texttt{new\_agent.fit} as \emph{False} if it is 0, an empty array, or any other \emph{False} value (such as \emph{None} or \emph{False}). The revised check is not only more concise but also more flexible, as it works with a broader range of \emph{False} values, ensuring that the assertion will hold even if \texttt{new\_agent.fit} is set to values other than 0 that are still considered \emph{False} in Python.

\begin{tcolorbox}
\textbf{RQ1 Summary:} The obtained results show GPT-4's effectiveness in refactoring anti-idiomatic code smells by applying established Pythonic idioms. It outperforms the benchmark in identifying additional refactoring opportunities while maintaining a high level of agreement on common refactorings, positioning GPT-4 as a competitive tool for enhancing Python code.
\end{tcolorbox}

\subsection{RQ2: GPT-4 Refactoring limitations}

Although rare, the \textit{bench\_more} metric highlights specific instances where the manually crafted benchmark outperforms GPT-4 by proposing a major number of refactoring for a given method, particularly in idioms such as \textit{ass\_multi\_tar}, where the benchmark identifies more refactoring opportunities in 18 methods.

Moreover, these cases where the benchmark suggests more refactorings underscore the complementary roles of GPT-4 and human-driven benchmarks. The benchmark, despite being manually crafted, may better account for specific coding standards or industry conventions that GPT-4 has not been explicitly trained on. As such, while GPT-4 can generate a broader set of refactorings, it may benefit from human insight in certain specialized contexts. These cases highlight the strength of having both automated and human-driven approaches in tandem, each covering the areas the other may miss.

From a practical standpoint, these findings suggest that GPT-4 and the manually developed benchmark can complement each other effectively. Developers could use GPT-4 for its ability to identify a wide range of refactoring opportunities and to automate routine tasks while relying on the benchmark for refining and validating those suggestions based on deeper, more contextual expertise. This hybrid approach would allow for a more thorough and nuanced refactoring process that combines the strengths of automated tools with human knowledge and insight.

These outcomes suggest that, despite GPT-4's superior performance overall, there are certain scenarios or idioms where the benchmark, created by developers with expert knowledge of Pythonic best practices, captures subtle patterns or nuances in code that GPT might miss. An example is shown in Listing~\ref{lst:less}, where a comparison between the refactoring proposed by GPT-4 and the refactoring proposed in the benchmark are displayed. In this case, the benchmark proposed three \textit{assign multiple target} refactorings, compared to the one proposed by GPT-4. The code is taken from the file \textit{test\_frame\_selector.py} from the \textit{OWOD} repository on GitHub~\cite{owod}.

To facilitate a clearer understanding of the refactorings, we have marked the variables that were originally defined separately and subsequently merged in the refactored versions. The refactorings proposed in the benchmark are as follows: (i) the first refactoring combines the definitions of the variables \texttt{\_SEED} and \texttt{\_K} into a single assignment expression, which is the same refactoring proposed by GPT-4; (ii) the second refactoring merges the assignments of the \texttt{selector} and \texttt{frame\_tss} variables; and (iii) the third and final refactoring combines the variables \texttt{\_SELECTED\_GT} and \texttt{selected} into one assignment. These changes contribute to a more compact and organized structure, where related variables are defined together in a manner that reflects their interdependence. 

Both the GPT-4 and benchmark refactorings align with Python’s best practices by adopting concise, idiomatic code that enhances readability and maintainability. The approach suggested by GPT-4 is a good starting point, offering a simple and effective refactoring of related variables. However, the benchmark goes a step further by applying the same idiomatic structure to additional variables, improving the overall clarity and compactness of the code.

 An aggressive refactoring can sometimes introduce subtle bugs or unintended consequences \cite{PentaBZ20}, especially in cases where variable assignments are not simply straightforward. GPT-4 might have recognized the simplicity and utility of merging \texttt{\_SEED} and \texttt{\_K} but avoided the other variables, possibly because they were more sensitive to context.

The \textit{gpt\_zero} column provides important insight into the coverage of GPT-4’s refactoring capabilities, indicating how frequently GPT-4 fails to suggest any refactorings for specific methods. With only 11 methods across all idioms receiving no refactorings, GPT exhibits broad applicability to identify potential improvements in Python code. This minimal number of zero-refactoring cases underscores GPT-4’s versatility and its general effectiveness in analyzing and suggesting refinements across a wide variety of Pythonic idioms.

However, the \textit{gpt\_zero} column also highlights certain limitations in GPT-4’s refactoring capabilities. For example, the idiom \textit{chain\_compare} has the highest \textit{gpt\_zero} count, with 6 methods identified as receiving no refactorings. This suggests that while GPT-4 surpasses in proposing improvements for many common Python idioms, it may encounter challenges when dealing with more complex or specific cases, such as those involving chained comparisons. The nature of such idioms might involve subtleties or patterns that GPT’s general model struggles to capture, potentially due to the particular syntactic structure or logic involved in the comparisons.

\begin{figure*}[t!]
\centering
\noindent\begin{minipage}{.45\textwidth}
\begin{lstlisting}
....
for pageno in intersection:
    my_page = self._pages[pageno]
    other_page = other._pages[pageno]
    if (my_page is None) ^ (other_page is None):
        changes[pageno] = None
    elif my_page is None:
        pass
    else:
        changed_offsets = my_page.changed_bytes(
            other_page, 
            page_addr=pageno * self.page_size)
        if changed_offsets:
            changes[pageno] = changed_offsets
....
\end{lstlisting}
\end{minipage}
\hspace{10pt}
\begin{minipage}{.45\textwidth}
\begin{lstlisting}
....
for pageno in intersection:
    my_page = self._pages[pageno]
    other_page = other._pages[pageno]
    if other_page is None is my_page:
        changes[pageno] = None
    elif my_page is None:
        pass
    else:
        changed_offsets = my_page.changed_bytes(
            other_page, 
            page_addr=pageno * self.page_size)
        if changed_offsets:
            changes[pageno] = changed_offsets
....
\end{lstlisting}
\end{minipage}
\captionof{lstlisting}{An example of refactoring where GPT-4 does not provide any action compared to the benchmark~\cite{zhang2024automated} solution. The Pythonic idiom used in this code is the \textit{chain comparison}. On the left side, the original code. On the right side, the refactoring proposed in the benchmark.}
\label{lst:zero}
\end{figure*}

\begin{table*}[t!]
    \caption{Correctness evaluation of refactorings generated by GPT-4 and the benchmark~\cite{zhang2024automated} for various Pythonic idioms. Metrics include the total refactorings proposed and incorrectly labeled refactorings for each approach}
    \begin{center}
        \begin{tabular}{|l|r|r|r|r|r|r|r|}
            \hline
            \textbf{Pythonic idiom} & \textbf{total} & \textbf{g\_ref} & \textbf{b\_ref} & \textbf{g\_correct} & \textbf{b\_correct} & \textbf{g\_wrong} & \textbf{b\_wrong}\\
            \hline
            list comprehensions & 10 & 24 & 11 & 21 & 7 & 3 & 4 \\
            \hline
            for multiple targets & 9 & 25 & 11 & 24 & 5 & 1 & 6 \\
            \hline
            chain comparison & 9 & 11 & 10 & 8 & 5 & 3 & 5 \\
            \hline
            fstring & 15 & 37 & 23 & 37 & 14 & 0 & 9 \\
            \hline
            assign multiple target & 10 & 20 & 16 & 19 & 14 & 1 & 2 \\
            \hline
             star in function calls & 8 & 10 & 10 & 9 & 5 & 1 & 5 \\
            \hline
            truth value test & 13 & 37 & 19 & 33 & 14 & 4 & 5 \\
            \hline
            loop else & 8 & 11 & 8 & 11 & 7 & 0 & 1 \\
            \hline
            with & 9 & 14 & 10 & 13 & 10 & 1 & 0 \\
            \hline
            dict comprehension & 8 & 14 & 8 & 10 & 7 & 4 & 1 \\
            \hline
            chain assign same value & 13 & 23 & 26 & 19 & 26 & 4 & 0 \\
            \hline
            set comprehension & 8 & 12 & 12 & 12 & 12 & 0 & 0 \\
            \hline
            \hline
            \textbf{total} & 120 & 238 & 164 & 216 & 126 & 22 & 38 \\
            \hline
        \end{tabular}
        \label{tab:correctness}
    \end{center}
\end{table*}

From a practical perspective, the relatively low number of \textit{gpt\_zero} instances indicates that developers can generally rely on GPT-4 for automated refactoring, but they may need to manually address edge cases or more complex patterns, like those found in \textit{chain\_compare}. This highlights an opportunity for further optimization of GPT-4’s refactoring capabilities, where a more targeted approach could help fill in the gaps for these specific cases.

Listing~\ref{lst:zero}  shows a comparison between the original code (on the left) and the refactoring proposed in the benchmark (on the right). In this case, the benchmark proposes a \textit{chain comparison} refactoring, whereas GPT-4 did not identify any applicable refactoring for this section of code. The code is taken from the file \textit{paged\_memory\_mixin.py} in the \textit{angr} repository on GitHub~\cite{angr}. For the sake of brevity and clarity, we present only a fragment of the code, focusing on the section where the refactoring occurs.

\begin{figure*}[t!]
\centering
\noindent\begin{minipage}{.45\textwidth}
\begin{lstlisting}
def main():
    print(""potential_field_planning start"")
    # Declaring variables
    sx = sy = gx = gy = grid_size = robot_radius 
        = None
    # Chained assignment
    sx, sy, gx, gy, grid_size, robot_radius = 
        0.0, 10.0, 30.0, 30.0, 0.5, 5.0  t
    ox = [15.0, 5.0, 20.0, 25.0]  
    oy = [25.0, 15.0, 26.0, 25.0]  
    if show_animation:
        plt.grid(True)
        plt.axis(""equal"")
    _, _ = potential_field_planning(
    sx, sy, gx, gy, ox, oy, grid_size, robot_radius)
    if show_animation:
        plt.show()
\end{lstlisting}
\end{minipage}
\hspace{10pt}
\begin{minipage}{.45\textwidth}
\begin{lstlisting}
def callback(self, chan, method_frame, 
    _header_frame, body, userdata=None):
        print(1)
        print(f'[x]{body}') # [x]%r'%body
\end{lstlisting}
\end{minipage}
\captionof{lstlisting}{Two examples of incorrect refactorings proposed by the two approaches. On the left side, the refactoring using \textit{chain comparison} idiom proposed by GPT-4. On the right side, the refactoring using \textit{fstring} idiom proposed by the benchmark~\cite{zhang2024automated}.}
\label{lst:correct}
\end{figure*}

The key modification proposed in the benchmark affects the fourth line of the code, specifically the if condition \texttt{if (my\_page is None) XOR (other\_page is None)}. The operator used in the code is the exclusive or (\emph{XOR}). This means that the condition will be true if exactly one of \texttt{my\_page} or \texttt{other\_page} is \emph{None}, but not both. Conversely, the refactoring condition applied in the benchmark is \texttt{if other\_page is None is my\_page}. This is a chained comparison, which checks if both conditions are true. This condition effectively translates to \texttt{if (other\_page is None) and (my\_page is None)}.
It is crucial to note that this refactoring introduces a logical alteration in the source code behavior. While the original \emph{XOR} condition checks for a situation where exactly one of the two variables is \emph{None}, the refactored chain comparison checks for the case where both are \emph{None}. This modification fundamentally changes the condition, potentially leading to unintended consequences depending on the intended functionality of the code. The original logic is designed to handle cases where one of the pages is absent (i.e., \emph{None}), while the refactored condition would only evaluate to true if both pages are absent.

In contrast, GPT-4 did not apply any refactoring to this part of the code. The model likely recognized the inherent risk in applying the chain comparison idiom here, as doing so would change the underlying logic of the condition. This cautious approach by GPT-4 highlights a critical aspect of refactoring: while some Pythonic idioms, such as chain comparisons, can improve readability and conciseness, their use can sometimes unintentionally modify the source code behavior. In this case, the refactoring proposed by the benchmark, while syntactically valid and adhering to Python conventions, changes the logic of the original condition and could result in unintended behavior in the context of the broader application.

\begin{tcolorbox}
\textbf{RQ2 Summary}: GPT-4 occasionally misses certain specialized cases. These findings highlight the complementary roles of GPT-4 and human-driven benchmarks, suggesting that an eclectic approach—leveraging the broad capabilities of GPT-4---alongside the expertise of manual benchmarks---can lead to a more comprehensive and context-sensitive refactoring process.
\end{tcolorbox}

\subsection{RQ3: Correctness evaluation}

Table~\ref{tab:correctness} presents the results of a correctness evaluation comparing the refactorings proposed by GPT-4 and the manually curated benchmark. For each idiom, the number of refactorings proposed by GPT-4 (\textit{g\_ref}) and the benchmark (\textit{b\_ref}) are displayed alongside their respective correctness metrics. Specifically, \textit{g\_correct} and \textit{b\_correct} represent the number of refactorings labeled as correct for GPT-4 and the benchmark, while \textit{g\_wrong} and \textit{b\_wrong} indicate the number of incorrect refactorings.

The table indicates that GPT-4 generally outperforms the benchmark in terms of correctness, with a total of 216 correct refactorings (90.7\% of GPT-4’s proposals) compared to 126 correct refactorings (76.8\% of the benchmark’s proposals). GPT-4 suggests more refactorings overall while maintaining a higher correctness rate across evaluated idioms.

For example, in the \textit{fstring} idiom, GPT-4 proposed 37 refactorings, with all 37 labeled as correct, showcasing its strong performance in this area. In contrast, the benchmark proposed 23 refactorings, with 14 deemed correct, which represents a lower success rate. Similarly, in the \textit{truth\_test} idiom, GPT-4 proposed 37 refactorings, of which 33 were correct (89\%), whereas the benchmark proposed 19 refactorings, with only 14 being correct (73.7\%).

While GPT-4 generally maintains higher correctness rates, the benchmark demonstrates notable performance in certain idioms. For instance, in the \textit{chain\_ass} idiom, the benchmark proposed 26 refactorings, all of which were correct, while GPT-4 only proposed 23 refactorings, with 19 correct. This example highlights that while GPT-4’s overall performance is superior, the benchmark can still sometime outperform GPT-4.

Additionally, there are cases where both methods suggest a similar number of refactorings with comparable correctness. For example, in the \textit{with} idiom, both GPT-4 and the benchmark proposed similar numbers of refactorings (14 for GPT-4 and 10 for the benchmark), with a nearly identical correctness rate—13 out of 14 for GPT-4 (92.9\%) and 10 out of 10 for the benchmark (100\%).

The number of incorrect refactorings (\textit{g\_wrong} and \textit{b\_wrong}) is relatively low for both methods, with GPT-4 producing 22 incorrect refactorings (9.2\% of its total) and the benchmark producing 38 incorrect refactorings (23.2\% of its total). These results suggest that while GPT-4 produces fewer incorrect refactorings, there are still areas where its suggestions can be improved, particularly for more complex idioms or cases requiring specialized knowledge.

The incorrect refactorings produced by the benchmark, while more frequent, may reflect the nature of the human-curated dataset, which is likely based on a fixed set of rules, heuristics, or predefined patterns. This human-driven approach, while ensuring a certain level of correctness and context-specific optimization, may not adapt as flexibly to a wide variety of coding styles or new refactoring opportunities that might emerge in novel programming contexts. This limitation may explain why the benchmark, despite its human insight, occasionally suggests errors more often than GPT-4.

Listing~\ref{lst:correct} shows two examples of incorrect refactorings proposed, on the left by GPT and on the right by the benchmark.

The GPT-proposed refactoring introduces a chain assignment \texttt{`sx = sy = gx = gy = grid\_size = robot\_radius = None`}. While this syntax is valid in Python, it is redundant and introduces unnecessary complexity. The code initializes multiple variables to None, but this step serves no practical purpose since the variables are immediately reassigned to new values shortly after \texttt{sx, sy, gx, gy, grid\_size, robot\_radius = 0.0, 10.0, 30.0, 30.0, 0.5, 5.0}. The original code, which directly assigns the new values to the variables without initializing them to None first, is cleaner and more readable. The introduction of None may create confusion, as a developer reading the code might wonder whether initializing the variables to None holds some special meaning or serves a specific purpose, but in this case, it does not. This unnecessary intermediate step adds cognitive overhead without offering any functional benefit. Furthermore, if this pattern were applied too widely, it could lead to situations where variables are initialized but never reassigned, making the code harder to debug and understand.

On the right side, the benchmark proposes a refactoring that alters how the body variable is printed. In the original code, the string formatting specifier \emph{\%r} is used: \texttt{print("\%r" \% body)}. The \emph{\%r} specifier calls the \texttt{repr()} function, which returns a string representation of an object that is often suitable for debugging or recreating the object. For example, if the body is a string like "Hello, World!", \texttt{repr('Hello, World!')} would return 'Hello, World!' with quotes, providing a clearer view of the actual value, including special characters. This is particularly useful when dealing with complex objects, as \texttt{repr()} provides more detailed and informative string representations compared to \texttt{str()}. However, the benchmark refactoring replaces \emph{\%r} with an \emph{f-string}: \texttt{print(f'[x]\{body\}')}. This \emph{f-string} uses the \emph{{body}} syntax, which calls \texttt{str(body)}, not \texttt{repr(body)}. The \texttt{str()} function is intended to return a human-readable string representation of the object, which may differ significantly from the \texttt{repr()} representation. For instance, \texttt{str('Hello, World!')} would output Hello, World! without quotes, whereas \texttt{repr('Hello, World!')} would return 'Hello, World!' with quotes. If the body is a complex object like a list or dictionary, the \emph{f-string} will likely produce a less informative output compared to the \texttt{repr(body)}, which recursively inspects objects and provides a more thorough representation. The decision to switch from \emph{\%r} to \emph{f-string} formatting could lead to potential issues, especially when debugging or logging complex objects, as the output might not provide enough detail to identify specific properties or differences in data.

\begin{tcolorbox}
\textbf{RQ3 Summary:} GPT-4 consistently outperforms the benchmark in terms of correctness, producing a higher percentage of correct refactorings. This highlights the potential of modern large language models (LLMs) in generating high-quality refactorings compared to deterministic, rule-based approaches.
\end{tcolorbox}

\section{Threats to Validity} \label{sec:threats}

\textbf{Construct Validity} addresses the alignment between theory and observation. The correctness evaluation of refactorings by GPT-4 and the benchmark was conducted manually, focusing on syntax and execution flow. Despite our expertise in Python code analysis, manual inspection is prone to error. Automating correctness validation by executing refactored code with test suites could improve reliability. However, existing project test suites may not cover the specific portions of code where refactorings are applied, requiring new test cases.

\textbf{Internal validity} pertains to the integrity of the study's execution. The non-deterministic nature of GPT-4, combined with the choice of hyperparameters and prompts, may influence the quality of the generated content without guaranteeing optimal solutions. Another potential threat is the random selection of refactorings during the correctness evaluation, which might introduce bias due to variations in project settings or the subset of selected code. To address this, we proportionally sampled 120 methods across different idioms, ensuring a representative distribution based on the prevalence of each idiom. Furthermore, since GPT-4’s training data is not publicly available, there is a possibility that similar code patterns exist in its training set, potentially providing an advantage over benchmark solutions. Finally, we used the default parameter of GPT-4o (temperature=1.0), which maximizes the variety of the model's response. However, our choice is because most developers use ChatGPT with default parameters since the UI does not provide settings to change the temperature. Keeping the default parameters, ensure that the experimental scenario try to be as much possible to a real scenario.

\textbf{External validity} concerns the generalizability of our findings. While our study focuses on twelve Pythonic idioms selected from the baseline dataset, this subset does not encompass all idioms in the PEP index, potentially limiting generalizability. However, prior studies emphasize these twelve idioms as among the most widely used and impactful in Python development, supporting their relevance. Additionally, our analysis involves methods extracted from diverse GitHub repositories, highlighting the practical need for addressing these anti-idiomatic patterns. The study focuses exclusively on GPT-4, without considering other LLMs. Nevertheless, GPT-4 is widely regarded as a state-of-the-art LLM, serving as a representative model. Future research could enhance generalizability by evaluating GPT-4 alongside other LLMs across a broader range of idioms.

\section{Conclusion} \label{sec:conclusion}

The adoption of idiomatic constructs in Python is widely recognized for enhancing code expressiveness, productivity, and efficiency, despite ongoing debates about their familiarity and understandability. This paper presented a replication study evaluating GPT-4's effectiveness in recommending idiomatic refactorings for Python code compared to a baseline benchmark. Our findings reveal that GPT-4 is highly effective in identifying and suggesting idiomatic constructs, often surpassing the baseline in scenarios where traditional methods fall short. However, in certain cases, GPT-4 fails to recommend appropriate idiomatic refactorings, highlighting areas for improvement. A manual evaluation of a random sample of recommendations confirmed the correctness and relevance of GPT-4's outputs, reinforcing its practical utility. These results suggest a complementary usage of GPT-4 and deterministic methods, paving the way for future research on hybrid solutions that combine the strengths of both approaches.

\section*{Acknowledgment}
We acknowledge the Italian ‘‘PRIN 2022’’ project TRex-SE: ‘‘Trustworthy Recommenders for Software Engineers’’, grant n. 2022LKJWHC.

\balance
\bibliographystyle{plain}
\bibliography{Biblio.bib}

\end{document}